\documentclass[preprint,showpacs,aps,prd,superscriptaddress,nofootinbib]{revtex4}

\usepackage{graphicx}
\usepackage{dcolumn}
\usepackage{epsfig}
\usepackage{color}
\usepackage{ulem}
\usepackage{amsmath}

\def\beq{\begin{equation}}
\def\eeq{\end{equation}}

\def\beqa{\begin{eqnarray}}
\def\eeqa{\end{eqnarray}}

\def\ket|#1>{\left|#1 \right>}

\def\bra<#1|{\left< #1 \right|}

\def\bracket<#1|#2>{\setbox0=\vbox{\hbox{$#1$$#2$}}\left<#1\kern1pt \vrule  height\ht0\kern2pt #2\right>}

\def\dirmat<#1|#2|#3>{\setbox0=\vbox{\hbox{$#1$$#2$$#3$}}\left<#1\kern1pt \vrule height\ht0\kern1pt#2\kern1pt \vrule height\ht0\kern1pt #3\right>}

\def\tev{{\rm TeV}}
\def\gev{{\rm GeV}}
\def\mev{{\rm MeV}}

\def\ePrime{\epsilon'}

\def\SM4{SM$_4$}
\def\CKM4{CKM$_4$}

\def\invfb{\ensuremath{{\rm fb}^{-1}}}

\def\Width#1{\ensuremath{\Gamma_{#1\rm -body}}}

\begin{document}

\begin{flushleft}
IMSc/2010/08/13 \\
UdeM-GPP-TH-10-192 \\
2918/10 \\
\end{flushleft}

\normalem

\title{
Measuring the Magnitude of the Fourth-Generation \CKM4 
Matrix Element $V_{t'b'}$
}

%%%%%%%%%%%%%%%%  %%%%%%%%%
\author{Diganta Das}
\affiliation{The Institute of Mathematical Sciences, Taramani, Chennai
  600113, India}

\author{David London}
\affiliation{Physique des Particules, Universit\'e de Montr\'eal,
  C.P. 6128, succursale centre-ville, Montr\'eal, QC, Canada H3C 3J7}

\author{Rahul Sinha}
\affiliation{The Institute of Mathematical Sciences, Taramani, Chennai
  600113, India}

\author{Abner Soffer} 
\affiliation{Tel Aviv University, Tel Aviv, 69978, Israel}

\date{\today}

\begin{abstract}
We study the decays of heavy chiral $(t', b')$ quarks in a
four-generation extension of the Standard Model. If the difference
between the $t'$ and $b'$ masses is smaller than the $W^\pm$ mass, as
favored by precision electroweak measurements, then the
Cabibbo-Kobayashi-Maskawa favored, on-shell decay $t' \to b' W^+$ is
forbidden. As a result, other $t'$ decays have substantial branching
fractions, which are highly sensitive to the magnitude of the diagonal
CKM matrix element $V_{t'b'}$. We show that $|V_{t'b'}|$ can be
determined from the ratio of the two-body and three-body $t'$-decay
branching fractions and estimate the precision of such a measurement
at the LHC. We discuss the hadronization of a $t'$ for large enough
values of $|V_{t'b'}|$.
\end{abstract}
\pacs{12.15.Ff, % Quark and lepton masses and mixing 
      12.15.Hh, % Determination of CKM matrix elements 
      14.40.Rt  % Exotic mesons
     }

\maketitle

\vskip .3 cm

%%%%%%%%%%%%%%%%%%%%
\section{Introduction}
\label{sec:intro}

One possible new-physics scenario that will be probed at the LHC is
the addition of a fourth generation to the standard model (SM),
resulting in a model generally denoted \SM4.  A fourth family brings
with it several interesting consequences~\cite{Holdom:2009rf}.  
It has implications for the status of and search for the Higgs
boson~\cite{Kribs:2007nz}, and can help remove the tension between the
SM fit and the lower bound on the Higgs-boson mass from LEP
\cite{Chanowitz:2009mz}. 
Adding only one extra generation to the SM is consistent with all
precision tests, even with a Higgs as heavy as
$500~\gev$~\cite{He:2001tp}. \SM4 is consistent with $SU(5)$
unification without supersymmetry~\cite{Hung:1997zj}.
Electroweak baryogenesis cannot be accomodated in the SM with three
generations, but it may be plausible in \SM4, due to the presence of
additional weak phases and the contribution of the heavy
fourth-generation quarks to the Jarlskog invariant in the
four-generation Cabibbo-Kobayashi-Maskawa matrix
(\CKM4)~\cite{Hou:2008xd}.
A rather heavy fourth generation could also naturally give rise to
dynamical electroweak-symmetry breaking~\cite{Bardeen:1989ds}, since a
quark with a mass of more than about half a~$\tev$ falls in the strong
coupling region~\cite{Chanowitz:1978mv}.

The CDF collaboration has searched for the decay of a heavy $t'$ quark
to a $W$ boson and a jet. Assuming this is the dominant decay mode,
they set a limit on the $t'$ mass, $m_{t'} > 256~\gev$ at 90\% C.L.
\cite{CDF:2008nf}. CDF has also looked for $b' \to t W^-$, setting the
limit $m_{b'}>338~\gev$ at 95\% C.L.~\cite{Aaltonen:2009nr}.
With higher energy and luminosity than the Tevatron, the LHC will have
much greater sensitivity for the discovery of fourth-generation quarks
and could measure some of their parameters.  Of particular importance
to flavor physics is the measurement of the \CKM4 matrix elements,
specifically $V_{t'b'}$, which is the main focus of this paper.

While a fourth generation is consistent with precision electroweak
data, stringent constraints exist on the deviation of the \CKM4
diagonal element $V_{t'b'}$ from unity, as well as on the mass
splitting of the fourth-generation quark doublet. In particular, the
best fit gives $\Delta m \equiv m_{t'} - m_{b'} \simeq 56~\gev$, where
we have assumed that the contribution from the fourth-generation
leptons is negligible due to a small lepton-neutrino mass
splitting.

Therefore, in this paper we explore the implications of \SM4 with
$\Delta m < m_W$, restricting the discussion to this scenario. The
decay $t' \to b' W$ is then not permitted, and one must consider other
decay modes. The three-body decay $t'\to b' W^{+*}$, where $W^*$ is an
off-shell $W$ boson producing a $f\bar f'$ pair, 
is kinematically suppressed. The decay involving
a virtual $b'$, $t'\to W {b'}^*$, is found to be even smaller. A large
value of $|V_{t'b'}|$ and the unitarity of \CKM4 lead to suppression of
two-body decays $t\to D W^+$, where $D\equiv \{d,s,b\}$.  Penguin
decays are both \CKM4- and loop-suppressed and are found to have negligible
branching fractions.
Calculating the rates of these processes, we show that for a
substantial range of the parameters, $|V_{t'b'}|$ can be determined from
the ratio of the branching fractions of $t'\to b' W^{+*}$
and $t\to D W^+$.

Now, it is well known that, because the $t$ quark decays rapidly,
there is not enough time for it to hadronize. On the other hand, we
find that, for a sizable part of the allowed parameter space, the
decay rates of the $t'$ are small enough that hadronization of this
fourth-generation quark can occur. The formation of such a bound state
is an interesting consequence of fourth-generation quarks.

This paper is organized as follows.
In Section ~\ref{sec:EW-constraints} we review constraints on the
fourth-generation parameters. We calculate the decay rates of
the $t'$ quark in Section~\ref{sec:BR-Gamma} for the parameter space
permitted by the electroweak constraints. 
We discuss the measurement of $V_{t'b'}$ and its uncertainties
in Section~\ref{sec:measuring}, 
and provide concluding remarks in Section~\ref{sec:summary}.

%%%%%%%%%%%%%%%%%%%%
\section{Electroweak Constraints}
\label{sec:EW-constraints}

The quantum oblique corrections~\cite{ref:oblique}, quantified in
terms of the $S$, $T$, and $U$ parameters, depend on the masses and
mixing parameters of the fourth-generation fermions. For $N_C=3$
colors, the contribution of the these fermions to $S$ and $T$
is\footnote{To a good approximation, NP that is heavy gives a negligible
  contribution to $U$.}
\cite{Chanowitz:2009mz}
\beqa
S_4&=& {1 \over 6\pi} 
     \left(3-\ln{\frac{x_{t'}}{{x}_{b'}}}\right) , \nonumber\\
T_4 &=& {3 \over 8 \pi s_W^2 (1-s_W^2)} \times \nonumber\\
&& \left[  |V_{t'b'}|^2 F_{t'b'}
           + \sum_{D=d}^{b} |V_{t'D}|^2 F_{t'D}
           + \sum_{U=u}^{t} |V_{Ub'}|^2 F_{Ub'}
           - |V_{tb}|^2 F_{tb}
	   + {F_{l_4 \nu_4} \over 3}
     \right] ,
\label{eq:ST}
\eeqa
where $s_W^2 \equiv \sin^2\theta_W$ and 
\beq
F_{ij} \equiv {x_i + x_j \over 2}
         -{x_i x_j \over x_i - x_j} \log{x_i \over x_j} ,
\eeq
with $x_i \equiv (m_i / m_W)^2$.
For approximately degenerate fermion masses, $x_i - x_j \ll x_j$,
$F_{ij} \sim (x_j - x_i)/2$, and the contribution to $T_4$ is small.
% In the other extreme, $x_i \gg x_j$, one finds $F_{ij} \sim x_i / 2$. 
In order to focus the discussion on the quark sector, in what follows
we take the fourth-generation neutrino and charged lepton to have very
similar masses, so that the contribution of $F_{l_4 \nu_4}$ to $T_4$
is negligible relative to that of the quarks.

In order to obtain the constraints from the oblique corrections, it is
necessary to take values for the \CKM4 matrix elements, in particular
$|V_{t'b'}|$. Limits on these matrix elements can be found by
considering the loop-level contributions of the fourth generation to
various flavor-physics observables. There have been several analyses
of \CKM4 in the past~\cite{CKM4past}. Recently, a complete fit has
been done~\cite{ADL}, and it was found that $V_{t'b'} = 0.998\pm
0.004$. This is to be compared with the oblique-parameter fits below.

\begin{figure}[!htbp]
 \begin{center}
 \includegraphics[width=0.5\textwidth]{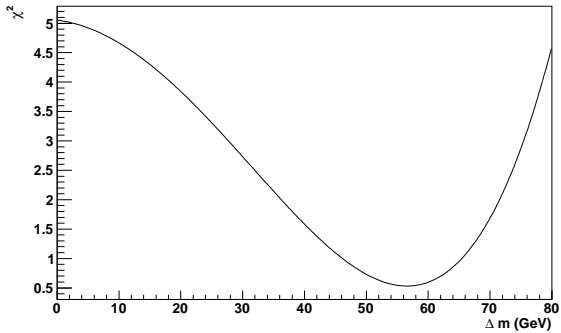} 
\caption{$\chi^2$ vs. the fourth-generation quark mass difference,
  given the $S$ and $T$ constraints in Ref.~\cite{ref:precisionEW},
  calculated with $U=0$, $m_t =171.4~\gev$, $m_H = 114~\gev$, and
  $m_{b'} = 400~\gev$.  }
\label{fig:chi2-vs-Dm}
\end{center}
\end{figure}

First, let us study the case $|V_{t'b'}| = 1$, for which only the
$F_{t'b'}$ term contributes to $T_4$. Taking $m_{b'} = 400~\gev$, 
We show in Fig.~\ref{fig:chi2-vs-Dm}, as a function of the
fourth-generation quark mass splitting $\Delta m \equiv m_{t'} -
m_{b'}$, the $\chi^2$ quantifying the agreement of $S_4$ and $T_4$
with the experimental constraints on $S$ and $T$, as given in
Ref.~\cite{ref:precisionEW} for $U=0$, $m_t =171.4~\gev$, and $m_H =
114~\gev$. One observes that the most likely value for $\Delta m$ is
around 56~\gev, while $\Delta m = 0$ and $\Delta m = m_W$ are
approximately equally disfavored at the 2.2 standard-deviation level.
A more general study~\cite{Erler:2010sk} that allows for
non-degenerate leptons finds $\Delta m$ to be even smaller.

\begin{figure}[!htbp]
 \begin{center}
\includegraphics[width=0.5\textwidth]{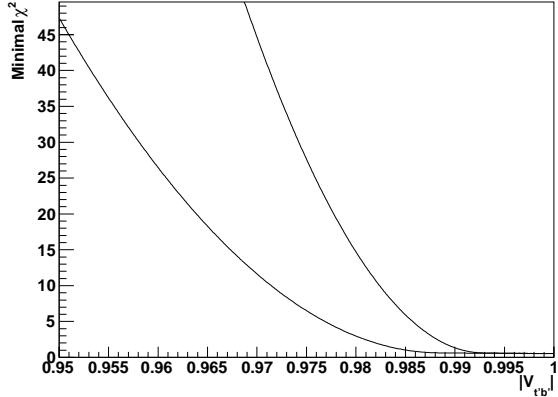} 
\caption{The minimal value of $\chi^2$ vs. $|V_{t'b'}|$ given $S$ and $T$ from
Ref.~\cite{ref:precisionEW}.  The lower (upper) curve is obtained when
the dominant off-diagonal CKM elements are $V_{t'b}$ and $V_{tb'}$
($V_{t'd}$ and $V_{ub'}$).}
\label{fig:minChi2-vs-Vt'b'}
\end{center}
\end{figure}

As one loosens the $|V_{t'b'}|$ requirement, the terms $F_{t'D}$ and
$F_{Ub'}$, which are of order $x_{t'}$ and $x_{b'}$ respectively, make
$T_4$ large and inconsistent with experimental results, given the
lower limits on the $t'$ and $b'$ masses. This puts a lower limit on
the value of $|V_{t'b'}|$ in a four-generation scenario. In
Fig.~\ref{fig:minChi2-vs-Vt'b'} we show the minimal $\chi^2$ obtained
for any $\Delta m > 0$ as a function of $|V_{t'b'}|$, from which
one can extract limits on $|V_{t'b'}|$. For example, requiring
$\chi^2<9$ yields $|V_{t'b'}|> 0.972$ ($|V_{t'b'}| > 0.983$) for the
case where the dominant off-diagonal CKM elements are $V_{t'b}$ and
$V_{tb'}$ ($V_{t'd}$ and $V_{ub'}$).

These results suggest that if there are four fermion families, then
\begin{itemize}

\item the CKM matrix remains highly diagonal, with $|V_{t'b'}| > 0.97$
  (this is consistent with, though weaker than, the result from the
  flavor-physics fit \cite{ADL}), and

\item the quark mass difference $\Delta m$ is smaller than the $W$ boson mass.

\end{itemize}
These two points will be assumed throughout the following sections.

%%%%%%%%%%%%%%%%%%%%
\section{The \boldmath $t'$ Decay Branching Fractions}
\label{sec:BR-Gamma}

With $\Delta m$ significantly smaller than $m_W$, the CKM-favored,
on-shell, two-body decay $t'\to b' W^+$ is kinematically forbidden.
The only processes leading to a $b'$ quark in the final state are the
off-shell decays $t'\to b'W^{+*}$ and $t'\to b'{^*} W^+$. Furthermore,
the amplitude for $t'\to b'{^*} W^+$ is additionally CKM suppressed at
the $b'{^*}$ decay vertex. Thus, the dominant decay 
with a $b'$ in the final state is the three-body process
$t'\to b'W^{+*}$. We find that the width for this decay is 
\beqa
\Width3 &\equiv& \nonumber\\
\Gamma\left(t'\rightarrow b' W^{+*}\right) &=& 
     {\alpha^2 \, |V_{t'b'}|^2 \, m_{t'}^5 \over 192\pi \, s_W^4 \, m_W^4}
  \Biggl\{r\epsilon \left(6r^2 - 3r \ePrime - \epsilon^2\right)
          + 3r^3 (r - \ePrime) \log(1 - \epsilon)
  \nonumber\\
    &&~~~~~~~~  +~{6r^3\left[2 - \ePrime(\epsilon + 2r) + r^2\right]
        \left[{\pi\over 2} 
           + \sin^{-1} 
             \left({\epsilon^2 - \ePrime r \over 2r\sqrt{1-\epsilon}}\right)
     \right] 
        \over \sqrt{4r - (\epsilon + r)^2}
     }
  \Biggr\} ,
\label{eq:Gamma3body-full}
\eeqa
where
\beqa
\epsilon &\equiv& 1 - {m_{b'}^2 \over m_{t'}^2} 
      \approx 2 \, {\Delta m \over m_{t'}}, \nonumber\\
\ePrime &\equiv& 2 - \epsilon, \nonumber\\
r &\equiv& {m_W^2 \over m_{t'}^2}.
\eeqa
%%%%%%%%%%%%
In our calculations we ignore the $W$ width. However, this changes the 
$W$ propagator by only 4\% for $\Delta m = 75~\gev$, and is hence
neglected.

\begin{figure}[!htbp]
 \begin{center}
\includegraphics[width=0.5\textwidth]{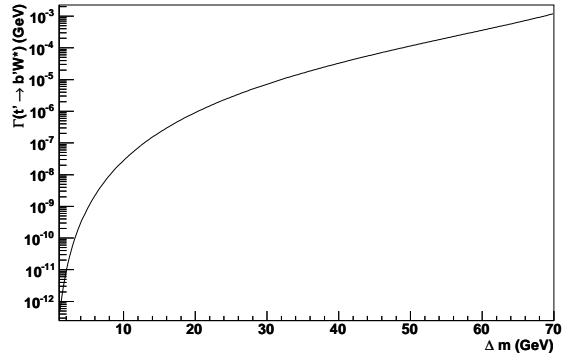} 
\caption{The decay rate $\Gamma\left(t'\rightarrow b' W^{+*}\right)$
from Eq.~(\ref{eq:Gamma3body-full}) for $m_{b'} = 400~\gev$.}
\label{fig:gamma-3body}
\end{center}
\end{figure}

In Fig.~\ref{fig:gamma-3body} we plot $\Gamma\left(t'\rightarrow b'
W^{+*}\right)$ as a function of $\Delta m$ for $m_{b'} = 400~\gev$ and
$|V_{t'b'}|=1$.  To understand the small magnitude of this decay
rate at small values of $\Delta m$, it is instructive to expand it 
in terms of $\epsilon$. The lowest-order term,
\beq
\Width3 \approx
{\alpha^2 \, |V_{t'b'}|^2 \, m^5_{t'} \over 960 \pi \, s_W^4 \, m_W^4} 
  \epsilon^5,
%%%%  \,  {\epsilon^5 \over \left(1 - \Delta m/m_W\right)^2} ,
\label{eq:3-body-1st}
\eeq
which differs from Eq.~(\ref{eq:Gamma3body-full}) by less than 30\%
for $\Delta m < 70~\gev$, demonstrates that
$\Gamma\left(t'\rightarrow b' W^{+*}\right)$ is suppressed by
$\epsilon^5$.

Due to this strong suppression of the CKM-favored $t'$ decay, the $t'$
may undergo CKM-suppressed decays with significant probability. The 
sum of the rates of the two-body decays $t'\to
D W^+$, where $D=d,s,b$, is 
\beqa
\Width2 &\equiv& 
\nonumber\\
 \sum_{D=\{d,s,b\}}\Gamma\left(t'\rightarrow DW^+\right) &=& 
 \frac{m_{t'}^3 \, \alpha}{16 \, s_W^2 \, m_W^2} \left(1-|V_{t'b'}|^{2}\right) 
     \left(1 - \frac{m_W^2}{m_{t'}^2}\right)^2
     \left(1 + 2 \, \frac{m_W^2}{m_{t'}^2}\right).  
\label{eq:2-body}
\eeqa
%%%%%%%%%%%%%%%%%%%%%%%%%%% 
This can be approximated as 
\beq
\Width2 \approx 
60 ~\gev \left({m_{t'} \over 455 ~\gev}\right)^3 \Delta V
\label{eq:2-body-approx}
\eeq
%%%%%%%%%%
where 
\beq
\Delta V \equiv 1 - |V_{t'b'}|
\eeq
%%%%%%%%%%%%%%
is much smaller than unity, and we have 
neglected terms of order $(m_W/m_{t'})^2$.

It is interesting to note that, unlike in the case of the top quark,
kinematic suppression of the CKM-favored $t'$ decay implies that for a
significant range of allowed values of $|V_{t'b'}|$, the $t'$ lifetime
is long enough for this quark to hadronize. Specifically,
Eq.~(\ref{eq:2-body-approx}) demonstrates that for 
$\Delta V < 3\times 10^{-3}$,
$\Gamma_{t'}$ is of order 100~\mev or less.  We
explore some consequences of hadronization in
Section~\ref{sec:measuring}.

For completeness, we also present the QCD-uncorrected
electroweak-penguin decay width~\cite{Burdman:1995te}:
\beq
\Gamma (t'\rightarrow U\gamma)=\frac{\alpha
  G_{F}^{2}}{128\pi^{4}} \, m_{t'}^{5}\sum_{Q=\{d,s,b,b'\}}
\left|V_{UQ}^{*}V_{t^{'}Q}F_{-{1\over 3}}\left({m_Q^2 \over m_W^2}\right)\right|^2,
\eeq
%%%%%%%%%%%%%%%%%%%%%%%%%%%%
where the factor
\beq
F_{Q_{em}}(x) \equiv
Q_{em}
\left[\frac{x^{3}-5x^{2}-2x}{4(x-1)^{3}}+\frac{3 x^{2}\ln x}{2(x-1)^{4}}\right]
+
\frac{2x^{3}+5x^{2}-x}{4(x-1)^{3}}-\frac{3x^{3}\ln x}{2(x-1)^{4}} 
\eeq
%%%%%%%%%%%%%%%%%
is most significant for large $x$, namely when the $b'$ is the internal
quark. In this case, we can write
\beq
\Gamma (t'\rightarrow U\gamma) \approx 1.6 \times 10^{-4}~\gev
   \left({m_{t'} \over 455~\gev}\right)^5 
    |V_{t'b'}|^2 \left(1 - |V_{t'b'}|^2\right),
\eeq
%%%%%%%%%%%%%%%%%%%%%%%%%%%%
which is several orders of magnitude smaller than 
the tree-level 2-body decay. 
We conclude that hadronic penguin decays are also unimportant, and 
disregard them in what follows.

%%%%%%%%%%%%%%%%%%%%%%%%%%%%%%%%%%%%%%%%%%%%%%%%%%%%%%%%%%%%%%%%%%%%%%%%%%%%%%%%
\section{Measuring \boldmath $|V_{t'b'}|$}
\label{sec:measuring}

Eqs.~(\ref{eq:Gamma3body-full}) and~(\ref{eq:2-body}) provide a
straightforward method to measure $|V_{t'b'}|$. We find that the ratio of
branching fractions of the two-body and three-body decays of the $t'$ is
\beq
R \equiv {\Width2  \over  \Width3}
  \propto {1-|V_{t'b'}|^2 \over |V_{t'b'}|^2},
\eeq
with a proportionality 
coefficient that depends on the $t'$ and $b'$ masses in a known way.
As we show below, this measurement is sensitive to very small deviations
of $|V_{t'b'}|$ from unity. In this case, $R\propto \Delta V$ to an
excellent approximation.
We therefore focus on $R$ as a measurement of the \SM4 parameter
$\Delta V$.

To estimate the statistical error of the $R$ measurement, we rely on
full-simulation studies performed for the LHC at $pp$ collision energies of 
14~\tev. Since the goal of those studies was to
estimate the discovery potential at low integrated luminosities, they 
made use of only the best decay channels and employ relatively
simple analysis techniques. It is anticipated that use of additional
channels, improved techniques, and more reliable understanding of the
detector and of 14~\tev\ processes will lead to reduced 
errors. However, we make no such assumptions in our estimates,
and simply extrapolate the sensitivities of these LHC studies
to the relevant parameter values.

For the two-body, $t' \to DW^+$ decay, we use an ATLAS
study~\cite{ATLAS-2body} performed for $m_{t'} = 500~\gev$.  The study
determined the discovery significance for this decay to be 9.2
standard deviations~($\sigma$) for an integrated luminosity of
1~\invfb. Only the $t' \bar t' \to l\nu+4$~jets channel was used, since it
provides an optimal combination of branching fraction and
signal-to-background ratio, making it most suitable for
low-luminosity discovery.

To estimate the statistical error on the three-body $t' \to b' W^{+*}$
signal yield, we use a CMS discovery study~\cite{CMS-3body} of $b'\bar b'$
production with $b'\to Wt$ decays. Such events are similar to the
events of interest to us, $t'\bar t'$, $t'\to b'W^{+*}$, $b'\to X$, in
that in both cases the final state contains at least ten jets, charged
leptons, and neutrinos. Therefore, we assume that the statistical
sensitivity determined in this study is a reasonable estimate for the
$t'\bar t'$, $t'\to b'W^{+*}$ case. The study found that a 400-\gev\ $b'$
can be discovered with $2.0\sigma$ significance given an integrated
luminosity of 0.1~\invfb, using the same-sign dilepton and trilepton
signatures.

\begin{figure}[!htbp]
 \begin{center}
\includegraphics[width=0.5\textwidth]{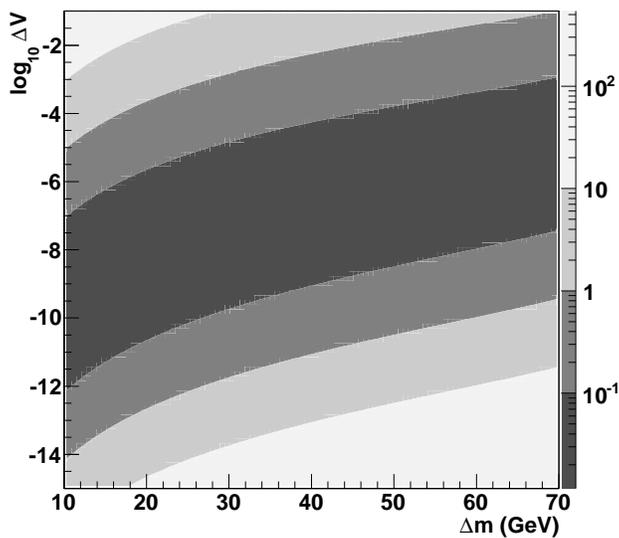} 
 \includegraphics[width=0.5\textwidth]{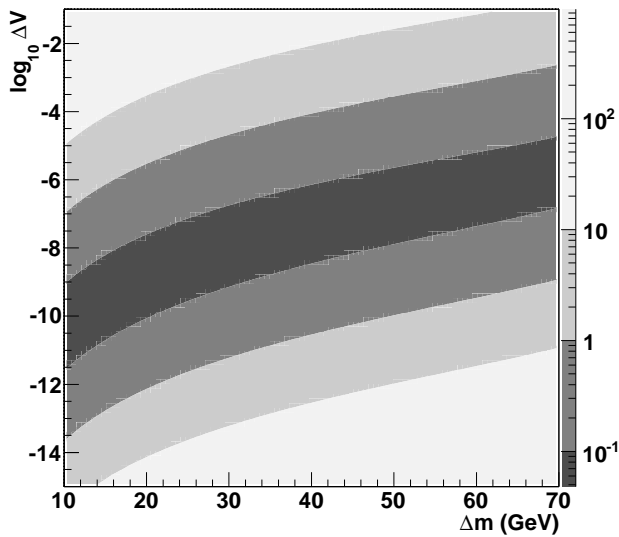}
\caption{Estimated relative statistical error 
$\sigma_{\Delta V}^{\rm stat} / \Delta V$ 
 as a function of $\log_{10} \Delta V$ and $\Delta m$, 
 assuming an integrated luminosity of 100~fb$^{-1}$ at one LHC experiment
 for (top) $m_{b'}=400~\gev$ or (bottom) $m_{b'}=700~\gev$.
 The decay $t'\to DW^{+}$ ($t'\to b'W^{+*}$) is the dominant $t'$
 decay channel above (below) the central dark band.}
\label{fig:stat-rel-error}
\end{center}
\end{figure}

We assume that these statistical errors scale with the square root of
the number of produced $t'\bar t'$ events. We take the integrated
luminosity to be 100~\invfb\ and obtain the cross section for different
$t'$ masses from PYTHIA~\cite{ref:pythia8}. 
Our results for $\sigma_{\Delta V}^{\rm stat} / \Delta V$, the
relative statistical error on $\Delta V$ from the measurement of the
ratio $R$, are shown in Fig.~\ref{fig:stat-rel-error} for $m_{b'} =
400~\gev$ and 700~\gev. The error is smallest when the statistical
uncertainties on the two branching fractions are equal.  These results
show that $\sigma_{\Delta V}^{\rm stat} / \Delta V$ smaller than 10\%
can be achieved over five (two) orders of magnitude of $\Delta V$
values for $m_{b'} = 400~\gev$ (700~\gev) and any given value of $\Delta m$.
%
%%%%%%%%%%%%%%%%%%%%%%%%
%%%%%%%%%%%%%%%%%%%%%%%%

Extraction of $\Delta V$ from $R$ using Eqs.~(\ref{eq:Gamma3body-full})
and~(\ref{eq:2-body}) requires knowledge of the masses of the $b'$ and
$t'$ quarks, introducing a systematic error $\sigma_{\Delta V}^{\rm syst}$, 
which we estimate next.

A general idea of the magnitude of the mass uncertainties may be
obtained from the projected LHC error on the top-quark mass.  
In an ATLAS Monte-Carlo study~\cite{ATLAS-tmass} for  
1~\invfb, a statistical error smaller than 0.4~\gev\ was estimated for $m_t$.
The systematic error was found to be about 1~\gev, if the
jet-energy-scale (JES) relative uncertainty, which is the largest
source of systematic error, is about 1\%.
The study reports that this JES uncertainty is obtainable with 1~\invfb\ for
light-quark jets, and we assume a similar error for
the $b$-quark jets.

To get an idea of the uncertainties on $m_{t'}$ and $m_{b'}$, we note
that while the cross section for $t'\bar t'$ production, at the masses
explored here, is two to three orders of magnitude smaller than for
$t\bar t$ production, this is mostly cancelled by the fact that the
integrated luminosity used in Ref.~\cite{ATLAS-tmass} is two orders of
magnitude smaller than our 100~\invfb benchmark. However, the detector
energy resolution generally degrades approximately as the square root
of the mass.  Therefore, given the errors quoted in
Ref.~\cite{ATLAS-tmass}, it is reasonable to conclude that the
fourth-generation quark masses may be known to order several~\gev.

\begin{figure}[!htbp]
 \begin{center}
 \includegraphics[width=0.5\textwidth]{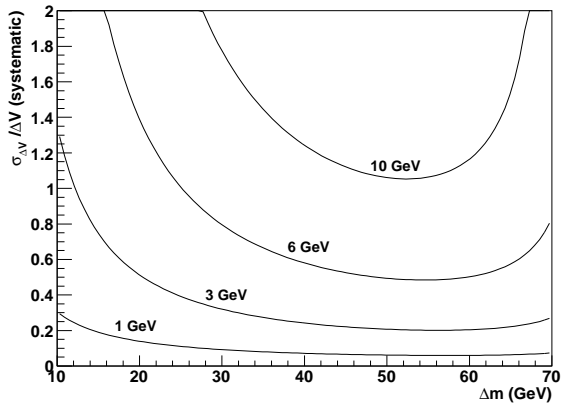}
\caption{Estimated relative systematic error $\sigma_{\Delta V}^{\rm
syst} / \Delta V$ due to the uncertainty on $\Delta m$, shown as a
function of $\Delta m$ for $\sigma_{\Delta m} = 1$, 3, 6, and 10~\gev.}
\label{fig:syst-rel-error}
\end{center}
\end{figure}

From Eqs.~(\ref{eq:3-body-1st}) and~(\ref{eq:2-body-approx}), we see
that to lowest order in $\Delta m$, one has the approximate
proportionality relation
\beq
{\Delta V \over R} \propto  {\Delta m^5 \over m_{t'}^3}.
\eeq
%%%%%%%%%%%%%%%%%
As a result, the impact of the $\Delta m$ uncertainty on the
relative error $\sigma_{\Delta V}^{\rm syst} /\Delta V$ is of order
$m_{t'} / \Delta m$ greater than that of the $m_{t'}$ uncertainty.
Therefore, we consider $\sigma_{\Delta m}$ as a main source
of systematic error.
Fig.~\ref{fig:syst-rel-error} shows the dependence of the
relative systematic error $\sigma_{\Delta V}^{\rm syst} / \Delta V$
on $\Delta m$ for several fixed values of $\sigma_{\Delta m}$.
%%%%%%%%%%%%%%%%%%%%%%%%
%%%%%%%%%%%%%%%%%%%%%%%%

We note that it may be possible to determine $\Delta m$ to better
precision than $m_{b'}$ or $m_{t'}$. For example, $\Delta m$ can be
determined from the spectrum of the lepton in the three-body decay $t'
\to b' W^{+*} \to b'l^+ \nu_l$.  Unless $\Delta m$ is large, this
lepton would be soft, distinguishing it from from the leptons arising
from the decays of the $b'$ and $t$ quarks in the event. Performing
this measurement requires the three-body branching fraction to be large,
which generally occurs for small values of $\Delta V$.
A detailed experimental feasibility study of such a measurement is
beyond the scope of this paper.

%%%%%%%%%%%%%%%%%%%%%%%%%%%%%%%%%%%%%%%
\subsection{\boldmath $t'$ Hadronization and $\Delta V$}

\begin{figure}[!htbp]
 \begin{center}
 \includegraphics[width=0.5\textwidth]{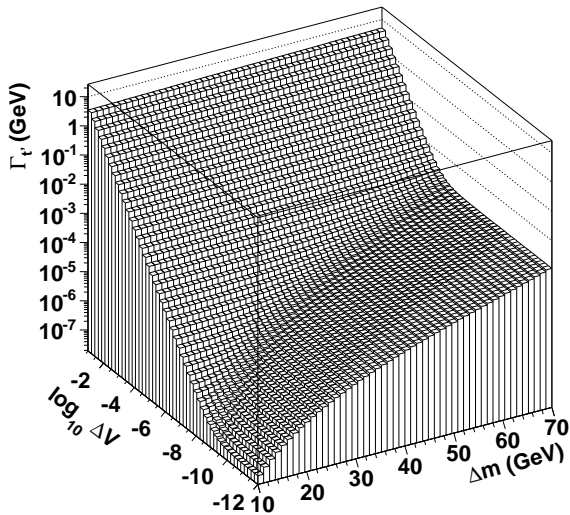} 
\caption{The total $t'$ width 
as a function of $\Delta m$ and $\Delta V$ for $m{b'} = 400~\gev$.}
\label{fig:Gamma}
\end{center} 
\end{figure}

As mentioned above, due to the kinematic suppression of the
CKM-favored transition, the $t'$ lifetime is large enough for it to
hadronize, if $\Delta V$ is smaller than about $3\times 10^{-3}$. In
principle, hadronization may be detected via depolarization of the
$t'\bar q$ meson, as outlined in
Ref.~\cite{Grossman:2008qh}. Depolarization occurs if $\Gamma_{t'} \ll
\Delta m_{T'}$, where $\Delta m_{T'}$ is the difference between the
masses of the spin-triplet vector meson and of the spin-singlet ground
state.  An estimate of this parameter is $\Delta m_{T'} \approx \Delta
m_B {m_B \over m_{T'}} \sim 1~\mev$.  From Fig.~\ref{fig:Gamma}, we
see that $\Gamma_{t'} < 0.1~\mev$ when $\Delta V < 10^{-6}$ and
$\Delta m < 45~\gev$. Thus, if one finds that $\Delta m < 45~\gev$,
searching for depolarization can help determine whether $\Delta V$ is
significantly above or below the $10^{-6}$ mark.

The information obtained from depolarization about $\Delta V$ is only
order-of-magnitude, and involves a complicated angular analysis that
can be performed only with the simple final state of the two-body $t'$
decay. This requires a large two-body branching fraction, and hence a
large value of $\Delta V$.  Given the $\Delta V < 10^{-6}$
depolarization condition, this implies that the depolarization
measurement of $\Delta V$ is practical only in regions where $\Delta V$ can
more easily be obtained from the measruement of the branching-fraction
ratio $R$.

%%%%%%%%%%%%%%%%%%%%%%%%
%%%%%%%%%%%%%%%%%%%%%%%% 

%%%%%%%%%%%%%%%%%%%%%%%%%%%%%%%%%%%%%%%%%%%%%%%%%%%%%%%%%%%%%%%%%%%%%
\section{Summary and Conclusions}
\label{sec:summary}

A fourth-generation extension of the Standard Model has several
attractive features, and studies show that it can be ruled out or
discovered with early LHC data.
Precision electroweak measurements imply that if fourth-generation
quarks exist, then the difference $\Delta m$ between their masses
is likely to be smaller than the $W$ boson mass.
If one assumes that this is indeed the case, then the two prominent $t'$ quark 
decays are the three-body, kinematically suppressed $t'\to b'W^{*+}$ and 
the two-body, \CKM4-suppressed $t'\to DW^+$.

We calculate the branching fractions of these decays and 
show that their ratio is a sensitive probe of $\Delta V$, the
difference of $|V_{t'b'}|$ from unity.
Using published simulation studies of fourth-generation quark
discovery at the LHC, we estimate the statistical error of this $\Delta
V$ measurement.
We estimate the main $\Delta V$ systematic error due to the
uncertainty on the mass difference $\Delta m$ and, for part of the
parameter space, suggest a possible way to obtain $\Delta m$ with
better precision than the individual quark masses.
An interesting consequence of the scenario described here is that the
$t'$ quark hadronizes if $\Delta V$ is of order $3\times 10^{-3}$ or
less. This is not likely to be useful for measuring $\Delta V$, but
could possibly be used as a laboratory for QCD in the infinite quark-mass
limit.

%%%%%%%%%%%%%%%%%%%%%%%%%%%%%%%%%%%%%%%%%%%%%%%%%%%%%%%%%%%%%%%%
\begin{acknowledgments}

We thank Shmuel Nussinov, Benjamin Svetitsky, and Ashutosh Kumar Alok
for valuable discussions and comments. The work of D.L. was
financially supported by NSERC of Canada.
\end{acknowledgments}
%%%%%%%%%%%%%%%%%%%%% REFERENCES %%%%%%%%%%%%%%%%%%%%%%%%%%%%%%%%

\end{document}